# Superconducting nanowire detector jitters limited by detector geometry

Niccolò Calandri[1,2], Qing-Yuan Zhao[1], Di Zhu[1], Andrew Dane[1], and Karl K.Berggren[1]

[1]*Department of Electrical Engineering and Computer Science, Massachusetts Institute of Technology, 77 Massachusetts Avenue, Cambridge, Massachusetts 02139, USA*

[2] *Department of Electronics, Information and Bioengineering (DEIB), Politecnico di Milano, via Leonardo da Vinci 32, 20133, ITA*

Detection jitter quantifies variance introduced by the detector in the determination of photon arrival time. It is a crucial performance parameter for systems using superconducting nanowire single photon detectors (SNSPDs). In this work, we have demonstrated that the detection timing jitter is limited in part by the spatial variation of photon detection events along the length of the wire. This distribution causes the generated electrical pulses to arrive at the readout at varied times. We define this jitter source as geometric jitter since it is related to the length and area of the SNSPD. To characterize the geometric jitter, we have constructed a novel differential cryogenic readout with less than 7 ps of electronic jitter that can amplify the pulses generated from the two ends of an SNSPD. By differencing the measured arrival times of the two electrical pulses, we were able to partially cancel out the difference of the propagation times and thus reduce the uncertainty of the photon arrival time. Our experimental data indicates that the variation of the differential propagation time was a few ps for a 3 μm × 3 μm device while it increased up to 50 ps for a 20 μm × 20 μm device. In a 20 μm × 20 μm large SNSPD, we achieved a 20% reduction in the overall detection timing jitter for detecting telecom-wavelength photons by using the differential cryogenic readout. The geometric jitter hypothesis was further confirmed by studying jitter in devices that consisted of long wires with 1-μm-long narrowed regions used for sensing photons.

Single-photon detectors are suitable for applications that require single-photon signal sensitivity from visible to infrared wavelengths. Quantum key distribution [1], time-of-flight laser ranging (LIDAR) [2], fluorescent imaging [3], and space based optical communication [4], all require single photon detectors. Superconducting nanowire single photon detectors (SNSPDs) excel relative to the other commercially available single-photon detectors [5], with efficiencies up to 96% at 1550 nm, response spectrum from 0.5 to 5 μm, low dark count rates below 100 count per second (cps), and a count rate above 100 million cps [7][8][9]. These applications require precise time resolution, and thus low timing jitter. Prior work has demonstrated that the jitter of the SNSPD can as low as 18 ps [11]. However, a wide range of jitter values—from 18 ps up to hundreds of ps—have been reported for SNSPDs of different sizes and different materials [7][8][8][9][11]. This inconsistency is an indication that the mechanisms involved in setting device jitter remain poorly understood.

The source of measured timing jitter can be separated into two main parts: (*i*) an electronic readout component, and (*ii*) an intrinsic component. The electronic component is due to noise from amplifiers, passive components, ground loops, and measurement equipment [10][11]. The existence of this limit is inferred from the observation that a reduction of the signal-to-noise ratio (SNR) and slew rate of the SNSPD signal increases jitter: with an RMS noise of 4 mV, a reduction in slew rate from 0.57 mV/ps to 0.15 mV/ps increased the electronic jitter from 6 ps to 24 ps [10]. The origin of the intrinsic jitter remains unclear: O'Connor [15] showed that a possible mechanism for intrinsic jitter is variation in the hotspot lengths. In this work, nanowire defects or constrictions distributed over the detector area were shown to produce different signal slopes and amplitudes, thus generating different SNSPD signals that increased jitter by up to 50 ps in a 20 μm × 20 μm device. However, limits of intrinsic jitter in devices without intrinsic inhomogeneity remains a puzzle.

Our recent discovery that a superconductive nanowire acts like a transmission line [12][13] has suggested a geometry-based approach to the understanding of intrinsic jitter. In particular, due to large kinetic inductivity of these wires, the signal velocity of such a transmission line is slower than the speed of light, while its characteristic impedance is much higher than 50 Ω. For a typical SNSPD the effective signal wavelength is thus reduced significantly, approaching the scale of the wire's physical length the wire's physical length. Therefore, instead of describing electrical pulse generation from the perspective of an SNSPD behaving like a lumped inductor, the device should be treated as a distributed circuit element, *i.e.* a transmission line. In the distributed-element scenario, after a photon is absorbed in the nanowire, two electrical pulses are generated which then propagate towards either end of the nanowire, arriving at times $t_1$ and $t_2$. Assuming the transmission line has a constant velocity $v$, $t_1$ and $t_2$ can be written as linear functions of the photon landing location $x_p$ and arrival time $t_p$, which are $t_1 = t_p + \frac{x_p}{v}$ and $t_2 = t_p + \frac{D-x_p}{v}$, where $D$ is the total length of the nanowire. In a typical SNSPD with a meandered shape, photons are usually uniformly illuminated on the detector. Consequently, the spatial variation of the absorbed photon will contribute to the variation of $t_1$ and $t_2$, introducing a geometric jitter $j_G$ that is closely related to the size of an SNSPD. Although this fully-distributed description oversimplifies the microwave behaviour of the SNSPD, it is qualitatively useful in understanding our results.

In this work, we have identified, studied, and provided experimental evidence for $j_G$ in an SNSPD by implementing a differential cryogenic readout to determine the absolute and relative arrival times of output pulses at the two ends of the nanowire. As shown in Fig. 1a, we pre-amplified these two pulses by means of two HEMT-based cryogenic amplifiers, increasing the signal-to-noise ratio (SNR) to reduce the electrical jitter $j_E$ below 7 ps. The amplifier had a high-input impedance and operated at 4.2 K while immersed in liquid helium (see Fig. 1a). The SNSPD was placed into a symmetric electrical network on a custom-printed circuit board (PCB), which was made from a 1.2-mm-thick FR4 laminate sheet. The SNSPD was biased with a DC current supplied by a voltage source with a serial resistor, while its output pulses were AC coupled to two HEMTs. [19]. In this way, positive and negative pulses arriving at each end of the SNSPD were separately amplified (Fig. 1b). The electrical noise was lowered by using multiple high-pass filters designed to have a low cut-off frequency around 200 MHz. The filter cut-off was defined by the input, output capacitors, and by the RC network placed at the source terminals of the HEMTs. Furthermore, the input stage of the amplifier consisted of a resistor and inductor; the SNSPD was shunted to the ground by a 50 Ω resistor, $R_{shunt}$ (that provided a low-impedance path, to avoid latching events) in series with a 10 μH conical broadband inductor $L_{shunt}$ (self-resonant frequency up to 40 GHz) used to boost the high-frequency component of the SNSPD signal. The conical inductor provided a high-impedance block for the rising edge of the SNSPD signal: in the frequency range between 200 MHz and 1.5 GHz (consistent with the bandwidth of the SNSPD signals), the input impedance of each HEMT was higher than 600 Ω. The output pulses were read out at room temperature, amplified by a 20MHz ~ 4-GHz-bandwidth amplifier and analysed with a 6 GHz oscilloscope. Fig 1b shows the output signals ($Ch_1$ and $Ch_2$) and the corresponding arrival times of the differential readout for a signal event from a 3 μm × 3 μm device of a nanowire meandered in a square geometry, biased at 24 μA. A clear delay was observed between $Ch_1$ and $Ch_2$. This delay varied due to variation in event position along the length of the wire. We achieved an output signal with an 800 mV amplitude, a rise time of 500 ps, and a slew rate of ~1 mV/ps. The RMS voltage output noise was lower than 5 mV, which guaranteed a maximum electronic RMS jitter of 5 ps (FWHM jitter of 8 ps) per channel [10].

We measured the pulse arrival times for both pulses to extract $t_1$ and $t_2$ after removing fixed delays from connections and amplifiers. As shown in Fig. 1c (histograms of $t_1$ and $t_2$) for the measurements of a 10 μm × 10 μm SNSPD, if we only use $t_1$ or $t_2$ to determine $t_p$, which is similar to readout of a conventional SNSPD, the overall detection timing jitter $j_{t1} = j_{t2} = 35$ ps, defined as the FWHM of the distribution of $t_1$ or $t_2$. This value includes not only the intrinsic jitter, but also our hypothesized geometric jitters $j_{G1}$ or $j_{G2}$ due to the distribution of photon landing locations along the wire length. For each photon-detection event, the difference of arrival times $\Delta t = t_2 - t_1 = \frac{(D-2x_p)}{2}$ is independent of photon arrival time $t_p$ and only depends on to the photon landing location $x_p$. Therefore, the variation of $\Delta t$, which we will call the differential jitter $j_\Delta$, is used as a metric to evaluate the contribution of the geometric jitter to the overall jitter. The sum of $t_1$ and $t_2$, $\Sigma t = t_2 + t_1 = 2t_p + \frac{D}{v}$ is a function of $t_p$ and independent of the photon landing locations $x_p$, from which $t_p$ can be determined by calculating $t_p = {\Sigma t - \frac{D}{v}}/{2}$, where $\frac{D}{v}$ is assumed to be constant. As shown in Fig. 1c (histogram of ($\frac{\Sigma t}{2}$)), the variation of $\frac{\Sigma t}{2}$ gives the detection sum timing jitter $j_\Sigma$, which is 29 ps, representing a 12% reduction relative to $j_{t1}$ or $j_{t2}$. A clear delay was observed between $t_1$ and $t_2$ for the hotspot nucleation event shown. This delay varied randomly from event to event, which we interpret as resulting from variation in position of the event along the length of the wire.

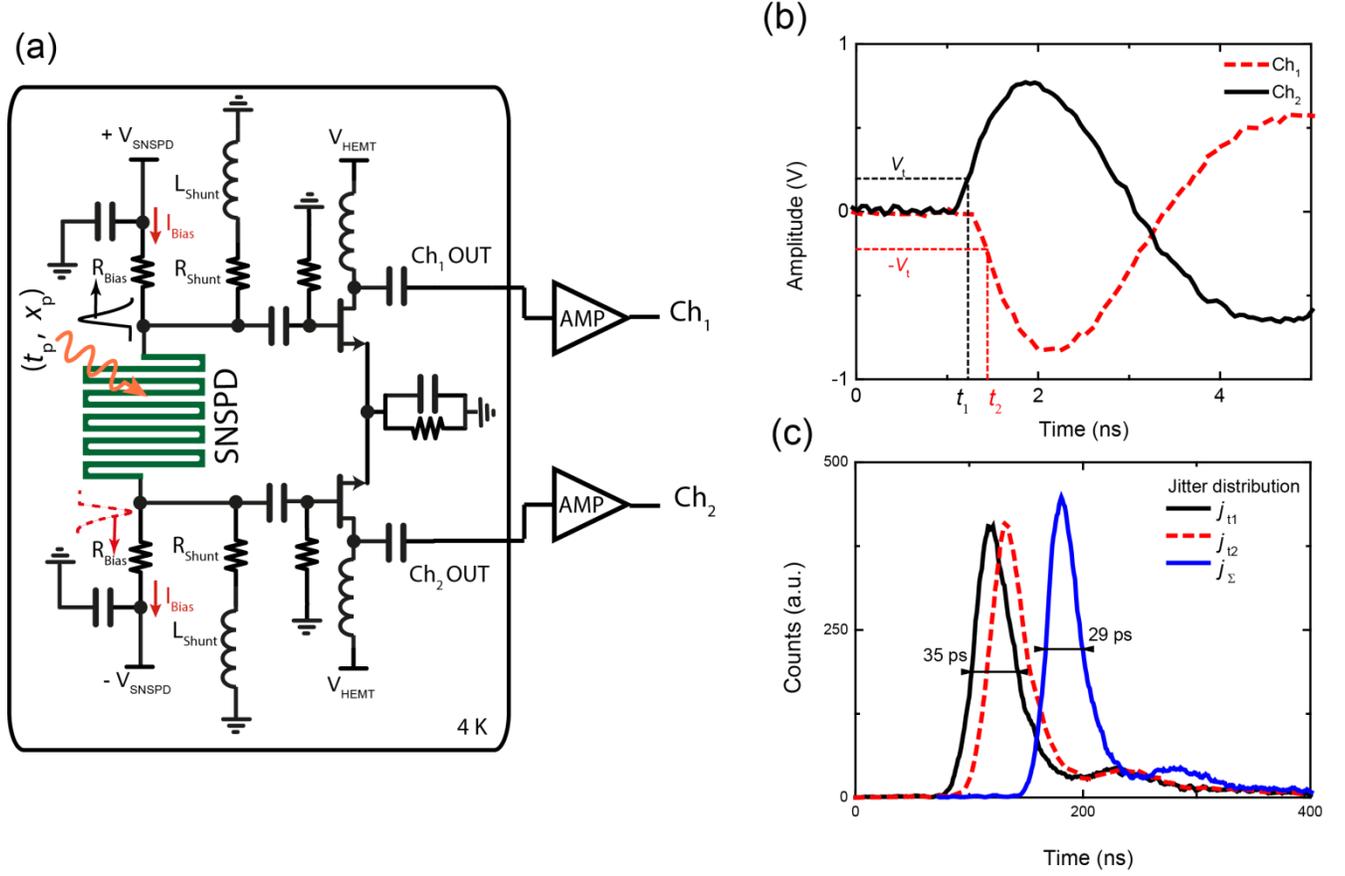

FIG. 1. (a) Schematic of differential cryogenic HEMT readout circuit with a NbN SNSPD. The HEMT amplifiers were designed to cut off all frequencies lower than 200 MHz to increase the SNR. The input impedance of the amplifier was set from the series combination of the shunt resistor and a conical inductor. All the components and the SNSPD were placed on a printed circuit board (PCB). (b) Output pulses of the differential readout acquired with a 6 GHz oscilloscope. The rising edge (10%–90% of the slope) is 500 ps wide and the slope is 1 mV/ps at 10% of the waveform amplitude. The delay time of the arrival signal from $Ch_1$ and $Ch_2$ is quoted at a defined threshold voltage $V_t$. (c) Jitter distribution $t_1$, $t_2$, and of the sum of two $t_1 + t_2 / 2$ of a 10 μm × 10 μm SNSPD.

With the metrics defined above, we implemented the differential readout on NbN SNSPDs of various geometries. The devices were fabricated from ~ 4-nm-thick NbN films on SiN$_x$-on-Si substrates. The standard meander structure was defined using electron-beam lithography, followed by reactive-ion etching. The nanowire width was 100 nm, the fill factor was 50%, and the detector area varied from 3 μm × 3 μm up to 20 μm × 20 μm. Assuming a uniform distribution in the hotspot stimulation from flood laser illumination on the active area, we measured $t_1$ and $t_2$ for each photon detection event and thus derived $j_\Delta$ and $j_\Sigma$.

Fig. 2a shows that increasing the dimension of the SNSPD increased the $j_\Delta$ as well. For larger SNSPDs, photons were spread over a larger area. Because it takes time for a pulse generated at one end to reach the other end, larger area resulted in a higher $\Delta t$. We compared the detection jitter $j_{t1}$ and $j_{t2}$, each measured from a single output, to the $j_\Sigma$ determined by using both outputs. As shown in Fig. 2c, as the size of the SNSPD was varied from 3 μm × 3 μm to 20 μm × 20 μm, jitters $j_{t1}$ and $j_{t2}$ increased from 22 ps to 45 ps. In comparison, the increase of $j_\Sigma$ over the same range was smaller, resulting in a reduction of 17 % in the variance of determination of photon arrival time by using $j_\Sigma$ relative to using $j_{t1}$ (or $j_{t2}$) for a 20 μm × 20 μm SNSPD.. As a result, we can use differential readout to reduce the single-photon detection jitter without changing the design of an SNSPD, and this reduction grows as the device area grows.

For sufficiently small detector areas, the geometric effect ought to be negligible, and so $j_\Delta$ should approach the intrinsic jitter of the hot-spot generation process while $j_\Sigma$ should approach $j_{t1}$ or $j_{t2}$. To test this hypothesis, we fabricated SNSPDs with a minimal detection area. For each detector, there was only a 1-μm-long and 100-nm-wide nanowire region for sensing photons. Because photon-detection events are limited to the 1-μm-long nanowire region, the geometric jitter from this small portion of the device $j_{\Delta S}$ should approach the limit set by the combined effect of variance in the intrinsic hot-spot generation process and electrical noise (as expected from [12] the electrical jitter increases with the overall inductance of the nanowire, including both narrower and wider regions). Hence, by comparing the difference between $j_\Delta$ in typical SNSPDs $j_{\Delta N}$ and $j_{\Delta S}$ in the 1-μm-long devices, as shown in Fig. 2b, we could evaluate how much geometric jitter $j_G$ contributes to the jitter of a typical device. after removing the contribution from electrical noise. We assumed that $j_{\Delta N}$ and $j_{\Delta S}$ add in quadrature, so $j_G = \sqrt{j_{\Delta N}^2 - j_{\Delta S}^2}$.

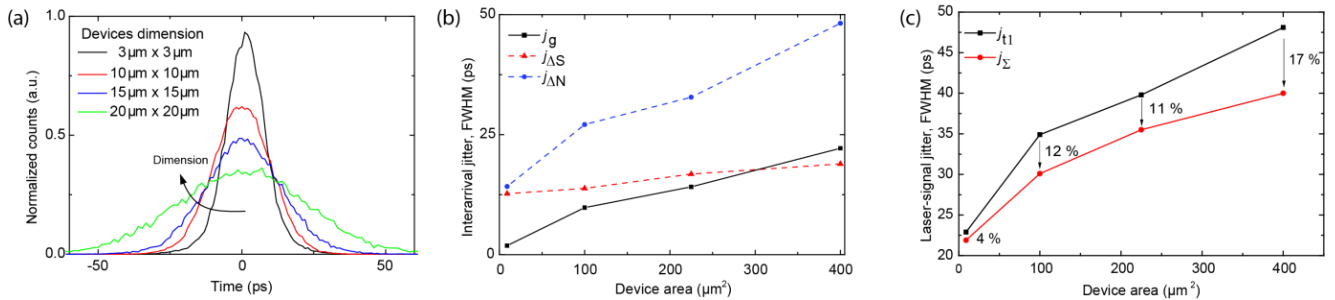

FIG. 2. (a) Distribution of $t_1$-$t_2$ for different detectors with different size: from 3 μm×3 μm up to 20 μm × 20 μm. The traces were acquired by using a 6 GHz oscilloscope, the start signal of the time correlated single-photon counting (TCSPC) was Ch$_1$, and Ch$_2$ was the stop. The devices were flood-illuminated by a 1550 nm picosecond pulsed laser, attenuated to the single photon level. (b) The full-width at half

maximum (FWHM) of $t_1$-$t_2$ for a single constriction device and for devices with varying area. The black line represents the final geometric jitter. (c) FWHM of SNSPD timing jitter acquired by using a 1550 nm picosecond pulsed laser, before and after propagation jitter correction for all the device geometries. The percentage improvement is indicated for each device.

Figure 3 shows the design of a comparison experiment, designed to characterize the maximum propagation time that can result when a photon hits at either edge of the nanowire. In this experiment, we used an on-chip multiplexing circuit to operate two SNSPDs (named L and R), which were designed to have limited photon-sensing area close to the edge of the meander shape shown in Fig. 3. Each of the SNSPDs had a 1-µm-long nanowire region placed in opposite corners, with the rest of the wire comprised of a 200-nm-wide nanowire. The devices had a total area of 20 µm × 20 µm of which only the 100 nm × 1 µm length was active. The arms were connected in parallel but oriented in opposite directions. The two SNPSDs were selectively biased by means of superconductive switches—nanocryotrons (nTrons) [16]—one placed in series with each side of the device (L and R) along the meander. When the nTron gate was biased over its critical current, the nTron channel became resistive and blocked bias current on one side of the device. Fig. 3FIG. **3**b shows the distribution of $\Delta t$ acquired in the two different configurations of Fig. 3a: (*i*) arm L ON, where the photon detection location is far from OUT$_1$; (*ii*) arm R ON, where the photon detection location was close to OUT$_2$. As shown in Fig. 3b, the two resulting distributions of $t_2$-$t_1$ had almost the same FWHM but had a spacing of $\Delta t$ = 42 ps between the distribution centres. $\Delta t/2$ thus correspond to the propagation delay for a signal to travel the entire length of the nanowire in this geometry (noting of course that the velocity of signal propagation along the 200 nm wide region will not be exactly the same as that for the 100-nm-wide region in typical devices).

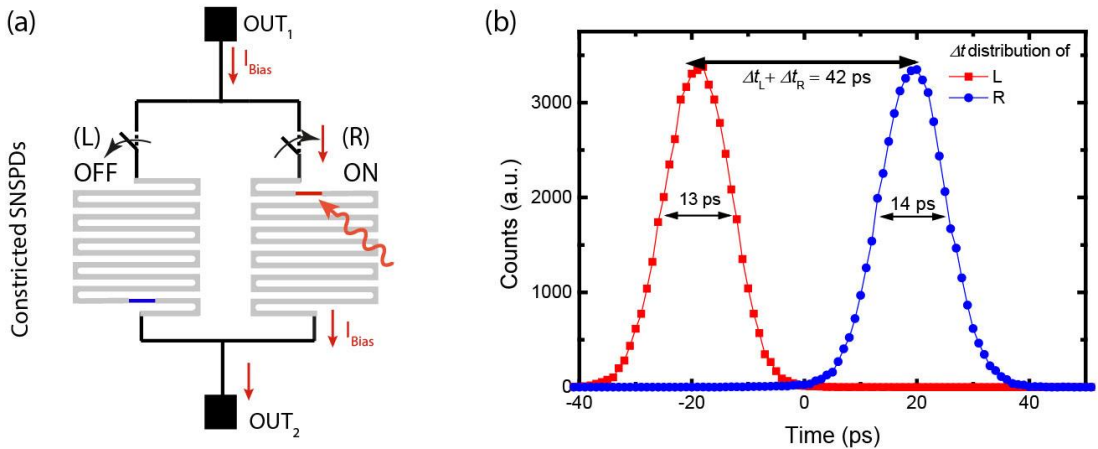

FIG. 3. (a) Schematic of the parallel single constriction devices. As depicted here, the R arm is biased and is able to detect photons, while L arm is OFF. The switches used were nTron devices: open when the gate (third terminal, not shown) was biased over its critical current, closed (i.e. zero resistance) when the gate was left floating. (b) **Δ*t*** distribution for the two sides of the double parallel device. The FWHM

of the distributions are quoted: $\Delta t_L$ and $\Delta t_R$ are the measured time differences between arrivals at channels 1 and 2 for the L and R devices respectively and correspond to the propagation delay for a signal to travel the entire length of the nanowire.

In conclusion we developed: (*i*) a differential cryogenic readout circuit demonstrating less than 7 ps of readout jitter by using an high impedance input stage; and (*ii*) a post-processing compensation method improving the detection timing jitter of an SNSPD, reducing the jitter by 17 % for a 20 μm × 20 μm device and thus achieving 38 ps of detection timing jitter. We also demonstrated the existence of geometric jitter in an SNSPD and characterized the dependence of geometric jitter on SNSPD size. We showed that this effect contributes more than 20 ps to the jitter of a 20 μm × 20 μm device. This work also implies that the observed system jitter should depend on how light is coupled to a detector (whether focussed on a single spot, or distributed evenly across the full device).

While this work has shown that some improvement in jitter is available by a simple change in readout method, we believe future work in which the full microwave environment and device characteristic is treated comprehensively could result in a substantial modification in the existing model of SNSPD operation. Novel SNSPD architectures can now be envisioned in which, for example, improved microwave biasing, readout, and signal amplification of the device is considered. We believe that any future efforts to drive SNSPD jitter and reset performance to higher levels will require consideration of the microwave characteristic of the nanowire geometry.

**Acknowledgements:**

This research was supported by the National Science Foundation (NSF) grant under contact NO. ECCS1-509486, and the Air Force Office of Scientific Research (AFOSR) grant under contract NO. FA9550-14-1-0052. Niccolò Calandri would like to thank his financial support from the Roberto Rocca project when he was a visiting student in MIT. Di Zhu is supported by National Science Scholarship from A*STAR, Singapore. Andrew Dane was supported by NASA Space Technology Research Fellowship, Grant # NNX14AL48H. Emily Toomey is acknowledged for helpful comments on the manuscript. Adam McCaughan is acknowledged for helpful discussions.